\documentclass[aps,floatfix]{revtex4}
\usepackage{graphics}
\usepackage{epsfig}
\usepackage{subfigure}
\usepackage{amsmath,amsfonts,amssymb,graphicx,times}

\begin{document}

\title{Immunization of traffic-driven epidemic spreading}

\author{Han-Xin Yang$^{1}$}\email{hxyang01@gmail.com}
\author{Bing-Hong Wang$^{2}$}

\affiliation{$^{1}$Department of Physics, Fuzhou University, Fuzhou
350108, China \\$^{2}$Department of Modern Physics, University of
Science and Technology of China, Hefei, 230026, China}

\begin{abstract}
In this paper, we study the control of the traffic-driven epidemic
spreading by immunization strategy. We consider the random,
degree-based and betweeness-based immunization strategies,
respectively. It is found that the betweeness-based immunization
strategy can most effectively prevent the outbreak of traffic-driven
epidemic. Besides, we find that the critical number of immune nodes
above which epidemic dies out is increased with the enhancement of
the spreading rate and the packet-generation rate.
\end{abstract}

\maketitle

Keywords: traffic-driven epidemic spreading; immunization strategy;
betweeness

\section{Introduction}

Both epidemic spreading~\cite{1,2,3,4,5,6,7,8,9,10,11} and traffic
transportation~\cite{12,13,14,15,16,17,18} on complex networks have
attracted much attention in the past decade. For a long time, the
two types of dynamical processes have been studied independently.
The first attempt to incorporate traffic into epidemic spreading is
based on metapopulation model~\cite{m1,m2,m3,m4,m5,m6}. This
framework describes a set of spatially structured interacting
subpopulations as a network, whose links denote the traveling path
of individuals across different subpopulations. Each subpopulation
consists of a large number of individuals. An infected individual
can infect other individuals in the same subpopulation. The
metapopulation model is often used to simulate the spread of human
and animal diseases (such as SARS and H1N1) among different cities.
In a recent work, Meloni $et$ $al.$ proposed another traffic-driven
epidemic spreading model~\cite{Meloni}, in which each node of a
network represents a router in the Internet and the epidemic can
spread between nodes by the transmission of packets. A susceptible
node will be infected with some probability every time it receives a
packet from an infected neighboring node. Meloni model can be
applied to study the propagation of computer virus in the Internet.

Meloni model has received increasing attention in recent years. It
has been found that the routing strategy plays an important role in
Meloni model. Epidemic spreading can be effectively controlled by a
local routing strategy~\cite{yang1} or an efficient routing
protocol~\cite{yang3}. For a given routing strategy, epidemic
spreading is affected by network structures. The increase of the
average network connectivity can slow down the epidemic
outbreak~\cite{yang4}. Besides, the epidemic threshold can be
enhanced by the targeted cutting of links among large-degree nodes
or edges with the largest algorithmic betweenness~\cite{yang5}.

Previous studies have shown that immunization is an effective way to
inhibit traditional epidemic spreading in which infections are
transmitted as a reaction process from nodes to all
neighbors~\cite{immune1,immune2}. In this paper, we will study how
different immunization strategies affect the traffic-driven epidemic
spreading in Meloni model. Three immunization strategies: the random
immunization, the targeted immunization of nodes with the largest
degree and the targeted immunization of nodes with the largest
algorithmic betweenness are considered, respectively. It is found
that, the targeted immunization of nodes with the largest
algorithmic betweenness can most effectively inhibit the
traffic-driven epidemic spreading.

\section{Traffic-driven epidemic spreading model and immunization strategies}\label{sec:model}

Following the work of Meloni $et$ $al.$~\cite{Meloni}, we
incorporate the traffic dynamics into the classical
susceptible-infected-susceptible model~\cite{SIS} of epidemic
spreading as follows. In a network of size $N$, at each time step,
$\lambda N$ new packets are generated with randomly chosen sources
and destinations (we call $\lambda$ as the packet-generation rate),
and each node can deliver at most $C$ packets towards their
destinations. Packets are forwarded according to a given routing
algorithm. The queue length of each agent is assumed to be
unlimited. The first-in-first-out principle applies to the queue.
Each newly generated packet is placed at the end of the queue of its
source node. Once a packet reaches its destination, it is removed
from the system. After a transient time, the total number of
delivered packets at each time will reach a steady value, then an
initial fraction of nodes $\rho_{0}$ is set to be infected (we set
$\rho_{0}=0.1$ in numerical experiments). The infection spreads in
the network through packet exchanges. Each susceptible node has the
probability $\beta$ of being infected every time it receives a
packet from an infected neighbor. The infected nodes recover at rate
$\mu$ (we set $\mu=1$ in this paper).

Once a node becomes immune, it cannot be infected and thus does not
transmit the infection to their neighbors. We consider three
immunization strategies respectively. (I) The random strategy (RS):
we randomly set $n$ nodes to be immunized from the network. (II) The
degree-based strategy (DS): we select $n$ nodes with the largest
degree to be immunized. (III) The betweenness-based strategy (BS):
we choose $n$ nodes with the largest algorithmic betweenness to be
immunized.

\section{Results and discussions}\label{sec:results}

\begin{figure}
\begin{center}
\scalebox{0.45}[0.45]{\includegraphics{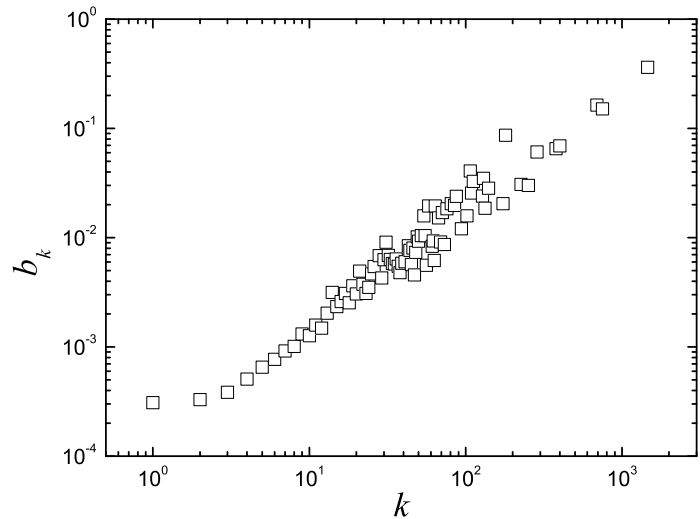}} \caption{The
dependence of the algorithmic betweenness $b_k$ on degree $k$. }
\label{fig1}
\end{center}
\end{figure}

In the following, we carry out simulations systematically by
employing traffic-driven epidemic spreading on top of the Internet
maps at the autonomous system level~\cite{real}, where the network
size $N = 6474$, the average degree $\langle k \rangle = 3.88$, and
the degree distribution follows a power law form $P(k) \sim k
^{-\gamma}$ with $\gamma \approx 2.2 $. Without special mention, we
use the the shortest-path routing algorithm to deliver packets.
Moreover, we assume that the node-delivering capacity $C$ is
infinite, so that traffic congestion will not occur in the network.

The algorithmic betweenness of a node $k$ is defined as~\cite{bc}:
\begin{equation}
b_{k}=\frac{1}{N(N-1)}\sum_{i\neq
j}\frac{\sigma_{ij}(k)}{\sigma_{ij}},
\end{equation}
where $\sigma_{ij}$ is the total number of possible paths going from
$i$ to $j$ according to a specific routing algorithm,
$\sigma_{ij}(k)$ is the number of such paths running through node
$k$, and the sum runs over all pairs of nodes. The algorithmic
betweenness of a node represents the average number of packets
passing through that node at each time step when the
packet-generation rate $\lambda=1/N$.

Figure~\ref{fig1} shows the dependence of the algorithmic
betweenness $b_k$ on degree $k$. A general trend is that, $b_k$
increases with $k$, and the relationship between $b_k$ and $k$
approximatively follows a power-law form as $b_k \sim k^{\nu}$. We
need to point out that, a larger-degree node may have lower
algorithmic betweenness. For example, $b_k$ is $8.7\times10^{-3}$
for $k=73$, which is smaller than that for $k=58$, whose algorithmic
betweenness is $2.0\times10^{-2}$ .

\begin{figure}
\begin{center}
\scalebox{0.45}[0.45]{\includegraphics{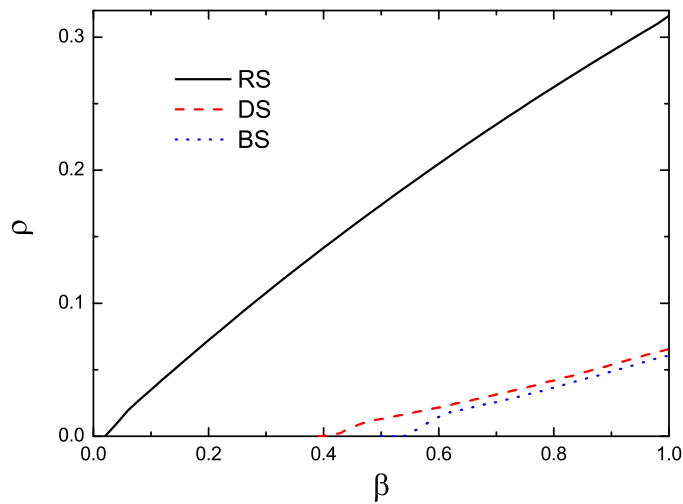}} \caption{ (Color
online) Density of infected nodes $\rho$ as a function of the
spreading rate $\beta$ in RS, DS, and BS cases. The
packet-generation rate $\lambda=0.3$ and the number of immune nodes
$n=50$. Each curve is an average of 100 different realizations.}
\label{fig2}
\end{center}
\end{figure}

Figure~\ref{fig2} shows the density of infected nodes $\rho$ as a
function of the spreading rate $\beta$ in RS, DS, and BS cases. One
can observe that for each case, there exists an epidemic threshold
$\beta_{c}$, beyond which the density of infected nodes is nonzero
and increases as $\beta$ is increased. For $\beta<\beta_{c}$, the
epidemic goes extinct and $\rho=0$.

\begin{figure}
\begin{center}
\scalebox{0.45}[0.45]{\includegraphics{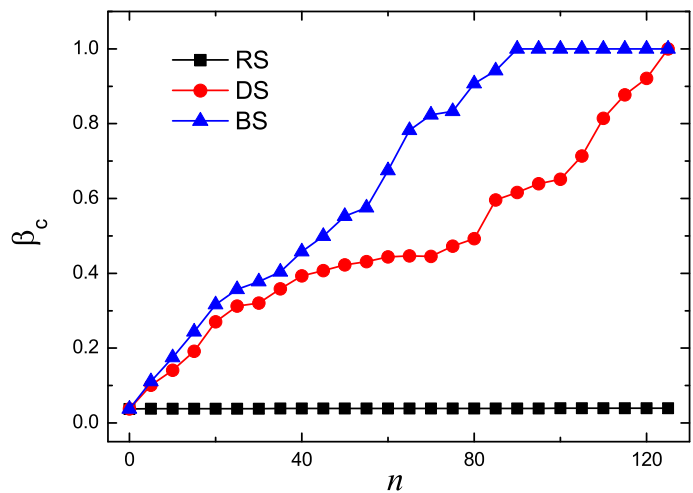}} \caption{(Color
online) The epidemic threshold $\beta_{c}$ as a function of the
number of immune nodes $n$ in DS and BS cases. The packet-generation
rate $\lambda=0.3$. Each data point results from an average over 100
different realizations.} \label{fig3}
\end{center}
\end{figure}

Figure~\ref{fig3} shows the epidemic threshold $\beta_{c}$ as a
function of the number of immune nodes $n$ in RS, DS, and BS cases.
For RS case, the epidemic threshold $\beta_{c}$ is almost unchanged
when the number of immune nodes is small, indicating that the random
immunization is useless. For DS and BS cases, $\beta_{c}$ increases
as $n$ increases, indicating that more immune nodes can better
suppress the outbreak of epidemic. Moreover, from Fig.~\ref{fig3},
one can find that for the same value of $n$, $\beta_{c}$ is the
highest in BS case. For BS case, only ninety immune nodes (about
1.4\% nodes) can completely prevent the outbreak of epidemic even
the spreading rate is 1. The above phenomena manifest that, BS is
the most effective immunization strategy in the traffic-driven
epidemic spreading.

\begin{figure}
\begin{center}
\scalebox{0.45}[0.45]{\includegraphics{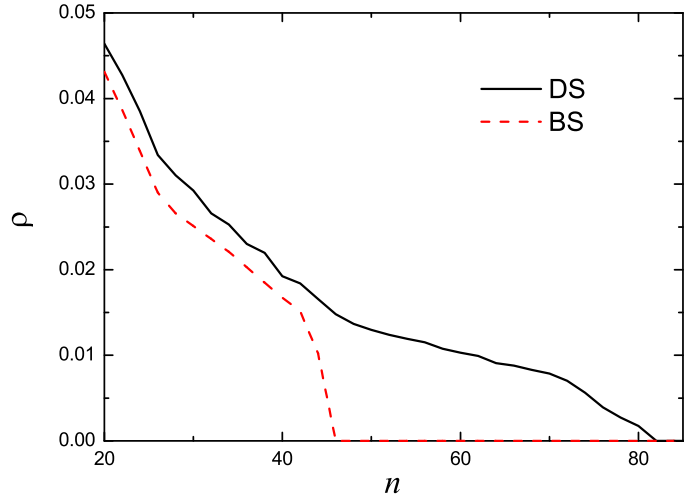}} \caption{(Color
online) The density of infected nodes $\rho$ as a function of the
number of immune nodes $n$ in DS and BS cases. The packet-generation
rate $\lambda=0.3$ and the spreading rate $\beta=0.5$. Each curve
results from an average over 100 different realizations.}
\label{fig4}
\end{center}
\end{figure}

\begin{figure}
\begin{center}
\scalebox{0.6}[0.6]{\includegraphics{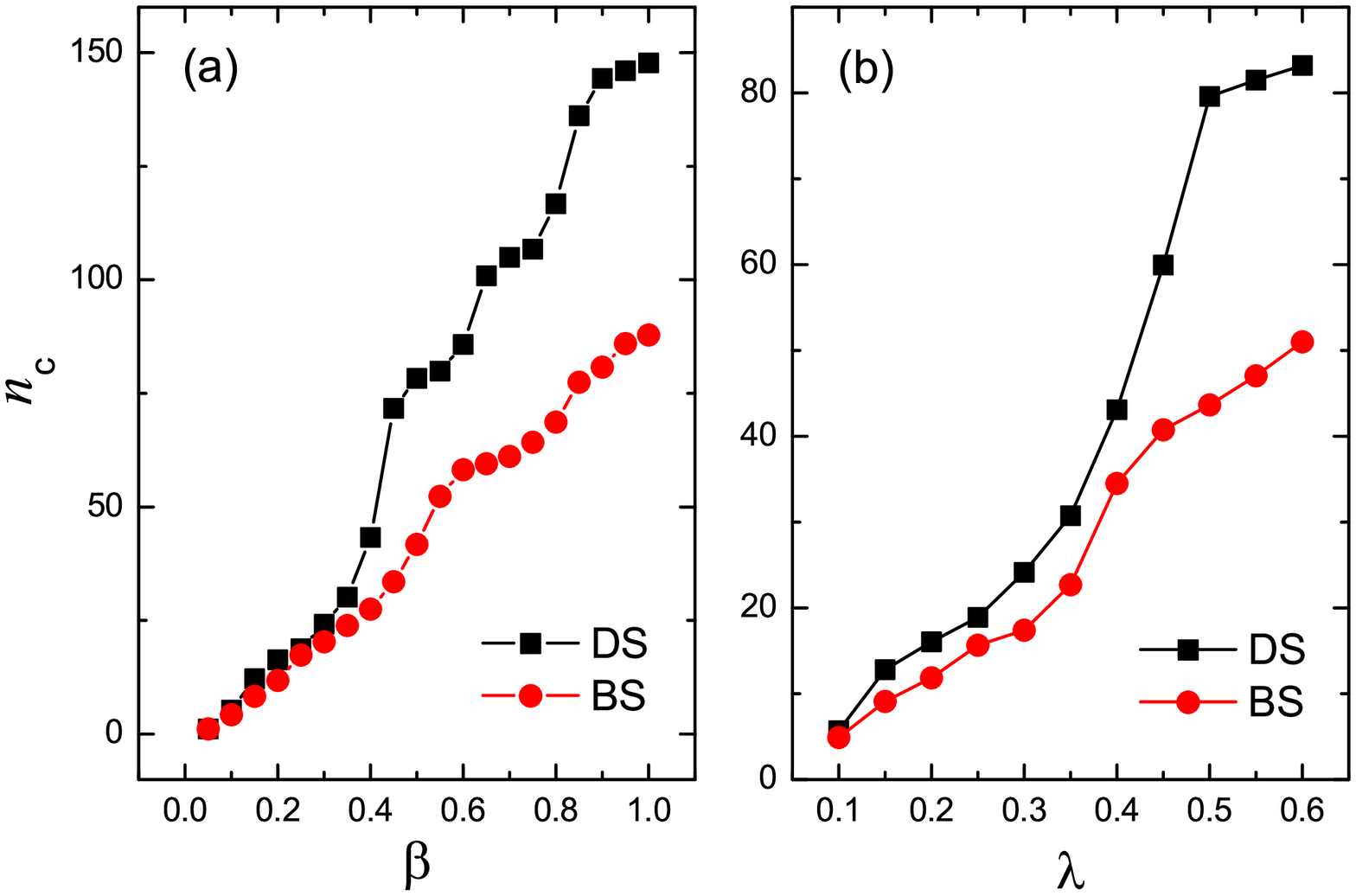}} \caption{(Color
online) (a) The critical number of immune nodes $n_{c}$ as a
function of the spreading rate $\beta$ in DS and BS cases. The
packet-generation rate $\lambda=0.3$. (b) The critical number of
immune nodes $n_{c}$ as a function of the packet-generation rate
$\lambda$ in DS and BS cases. The spreading rate $\beta=0.3$. Each
data point results from an average over 100 different realizations.
}\label{fig5}
\end{center}
\end{figure}

Figure~\ref{fig4} shows the density of infected nodes $\rho$ as a
function of the number of immune nodes $n$ in DS and BS cases. One
can observe that there exists a critical number of immune nodes
$n_{c}$, above which the epidemic goes extinct and $\rho=0$.
Undoubtedly, the smaller $n_{c}$ reduces the cost of immunization.
Figure~\ref{fig5}(a) shows that $n_{c}$ as a function of the
spreading rate $\beta$ in DS and BS cases. One can see that $n_{c}$
increases with $\beta$. Besides, for the same value of $\beta$,
$n_{c}$ is smaller in the case of BS than that of DS.
Figure~\ref{fig5}(b) shows that $n_{c}$ as a function of the
packet-generation rate $\lambda$ in DS and BS cases. One can see
that $n_{c}$ increases with $\lambda$, manifesting that the increase
of traffic flow is unfavorable for the control of epidemic
spreading. For the same value of $n$, $n_{c}$ is smaller in the case
of BS than that of DS. From Fig.~\ref{fig5}, one can find that,
compared to the degree-based immunization strategy, the
betweenness-based immunization strategy is more effective in
suppressing the traffic-driven epidemic spreading.

\begin{figure}
\begin{center}
\scalebox{0.45}[0.45]{\includegraphics{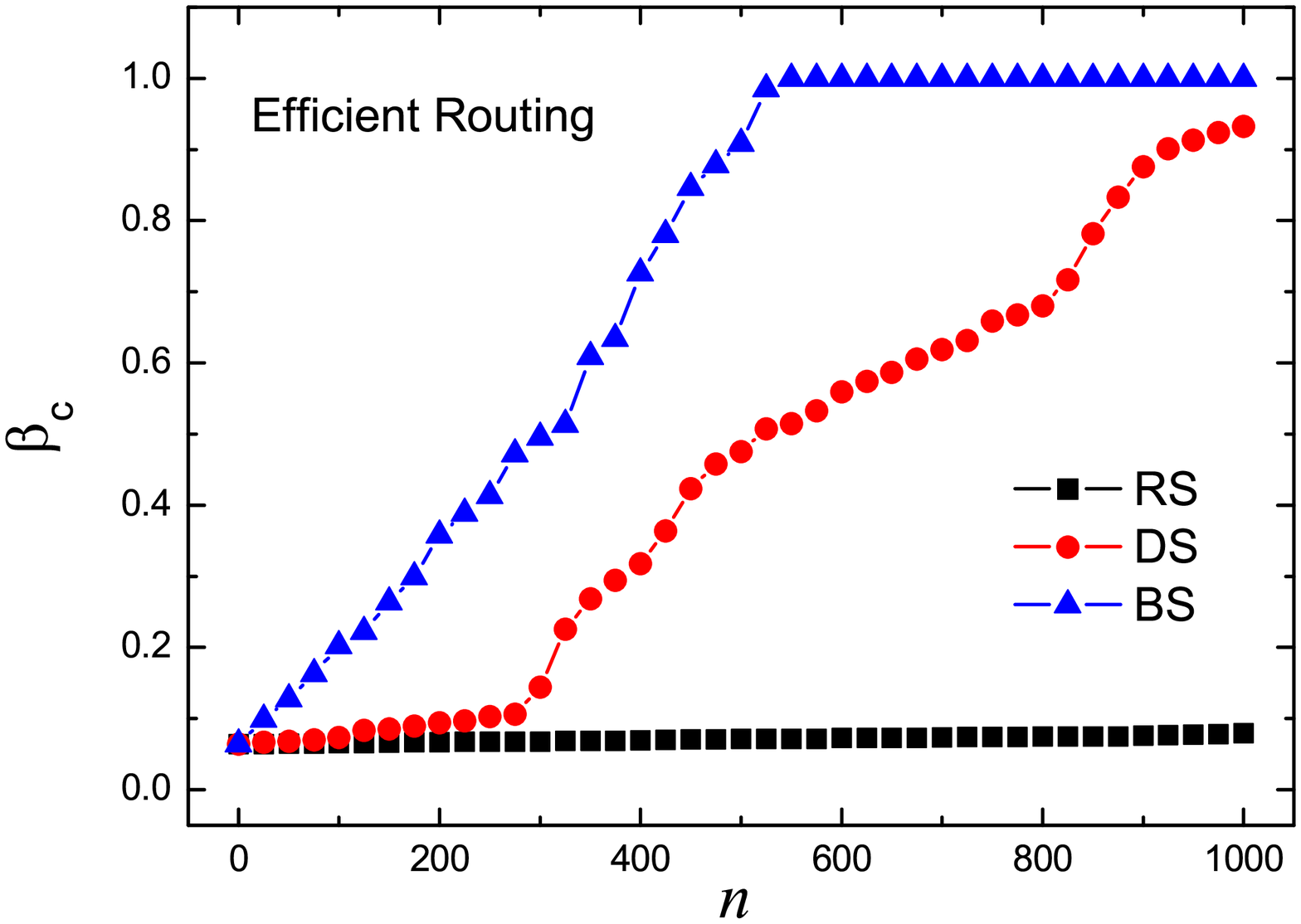}} \caption{(Color
online) The epidemic threshold $\beta_{c}$ as a function of the
number of immune nodes $n$ in DS and BS cases when the efficient
routing algorithm is applied. The packet-generation rate
$\lambda=0.3$. Each data point results from an average over 100
different realizations.} \label{fig6}
\end{center}
\end{figure}

In all the above studies, packets are forwarded following the
shortest-path routing algorithm. In fact, our finding that
betweenness-based immunization strategy can most effectively prevent
the outbreak of traffic-driven epidemic is robust with respect to
different kinds of routing algorithms. Figure~\ref{fig6} shows the
epidemic threshold $\beta_{c}$ as a function of the number of immune
nodes $n$ in RS, DS, and BS cases when the efficient routing
algorithm is applied. The efficient routing algorithm is described
as follows~\cite{ef}. For any path between nodes $i$ and $j$, $P(i
\rightarrow j) : = i \equiv x_{1}, x_{2}, \cdots , x_{n}\equiv j$,
we define
\begin{equation}\label{Eq2}
L\left(P(i \rightarrow
j):\alpha\right)=\sum_{l=1}^{n}k(x_{l})^{\alpha},
\end{equation}
where $k(x_{l})$ is the degree of node $x_{l}$ and $\alpha$ is a
tunable parameter. For any given $\alpha$, the efficient path
between $i$ and $j$ is corresponding to the route that makes the sum
$L\left(P(i \rightarrow j):\alpha\right)$ minimum. In this paper, we
set the routing parameter $\alpha=1$. From Fig.~\ref{fig6}, one can
find that for the same number of immune nodes $n$, $\beta_{c}$ is
the largest in BS case while $\beta_{c}$ keeps almost unchanged in
RS case.

\section{Conclusion}\label{sec:conclusion}

In conclusion, we have studied the effects of immunization
strategies on traffic-driven epidemic spreading. Our results show
that, the outbreak of traffic-driven epidemic can be effectively
suppressed when a small fraction of nodes with the largest
algorithmic betweenness are immunized. This finding is robust with
respect to different kinds of routing algorithms including the
shortest-path routing algorithm and the efficient routing algorithm.
For traditional epidemic spreading where infections are transmitted
as a reaction process from nodes to all neighbors, an effective
immunization strategy is to vaccinate the largest-degree
nodes~\cite{immune1}. However, compared to the betweenness-based
immunization strategy, the degree-based immunization strategy is
less efficient in the suppression of traffic-driven epidemic
spreading. This is because the larger-degree nodes may not have
higher algorithmic betweenness. Moreover, we find that more immune
nodes are needed to prevent the outbreak of epidemic as the
spreading rate and the packet-generation rate are increased. We hope
our results can be useful to control traffic-driven epidemic
spreading.

\begin{acknowledgments}
This work was supported by the National Science Foundation of China
(Grant Nos. 61403083, 11275186, 91024026 and 71301028), and the
Natural Science Foundation of Fujian Province, China (Grant No.
2013J05007).
\end{acknowledgments}

\end{document}